\documentclass[english]{paper}
\usepackage{color}
\usepackage[english]{babel}                                                          
\usepackage{amssymb}
\usepackage{psfrag}
\usepackage{epsfig}                                                          
\begin{document}                                                             
\newcommand{\be}{\begin{equation}}
\newcommand{\ee}{\end{equation}}
\newcommand{\bestar}{\[}
\newcommand{\eestar}{\]}
\newcommand{\beastar}{\begin{eqnarray*}}
\newcommand{\eeastar}{\end{eqnarray*}}
\newcommand{\beq}{\begin{equation}}
\newcommand{\eeq}{\end{equation}}
\newcommand{\bea}{\begin{eqnarray}}
\newcommand{\eea}{\end{eqnarray}}
\newcommand{\dfrac}{\displaystyle\frac}                                    
\newcommand{\disp}{\displaystyle}                                    
\newcommand{\mbf}{\mathbf}
\newcommand{\dint}{\displaystyle\int}
\renewcommand{\u}{\underline}                                        
\renewcommand{\o}{\overline}                                        
\newcommand{\ei}{\end{itemize}}
\newcommand{\bfig}{\begin{figure}[htb]\begin{center}}
\newcommand{\efig}{\end{center}\end{figure}}
\newcommand{\eqm}[2]{~(\ref{#1}-\ref{#2})}
\newcommand{\eq}[1]{~(\ref{#1})}
\newcommand{\eqq}[2]{~(\ref{#1},\ref{#2})}
\newcommand{\eqqq}[3]{~(\ref{#1},\ref{#2},\ref{#3})}
\newcommand{\order}{{{\mathcal O}}}
\newcommand{\ie}{{\it i.e.}}
\newcommand{\eg}{{\it e.g.}}
\newtheorem{thm}{Theorem}     
\newtheorem{obs}{Remark}  
\newcommand{\f}{^{\rm f}}
\newcommand{\G}{^{\rm G}}
\renewcommand{\L}{^{\rm L}}
\newcommand{\g}{^{\rm g}}
\renewcommand{\l}{^{\rm l}}
\newcommand{\W}{^{\rm W}}
\newcommand{\ph}{^a}
\newcommand{\parent}{_0}
\newcommand{\tot}{_{\rm{tot}}}
\newcommand{\src}{\psi}
\newcommand{\out}{^{\rm out}}
\newcommand{\vap}{^{\rm vap}}
\newcommand{\cond}{^{\rm cond}}
\newcommand{\esp}{_{\rm exp}}
\newcommand{\LG}{_{\rm LG}}
\newcommand{\WG}{^{\rm WG}}
\newcommand{\WL}{^{\rm WL}}
\newcommand{\mol}{n}
\newcommand{\molfrac}{x}
\newcommand{\vol}{v}
\newcommand{\Vol}{V}
\newcommand{\vel}{w}
\newcommand{\surf}{A}
\newcommand{\Ent}{H}
\newcommand{\fen}{f}
\newcommand{\Fen}{F}
\newcommand{\Fid}{F_{\rm id}}
\newcommand{\Fexc}{\widetilde F}
\newcommand{\Gibbs}{G}
\newcommand{\gibbs}{g}
\newcommand{\entr}{s}
\newcommand{\Entr}{S}
\newcommand{\Inte}{U}
\newcommand{\ent}{h}
\newcommand{\inte}{u}
\newcommand{\temp}{T}
\newcommand{\pres}{P}
\newcommand{\heat}{Q}
\newcommand{\molheat}{q}
\newcommand{\comp}{z}
\newcommand{\calv}{c_\vol}
\newcommand{\calp}{c_\pres}
\newcommand{\crit}{_{\rm c}}
\newcommand{\hcoeff}{\lambda}
\newcommand{\noz}{_{\rm noz}}
\newcommand{\text}{\mbox}
\date{}
\title{Blowdown of hydrocarbons pressure vessel with partial phase separation}
\author{A. Speranza\\
Industrial Innovation Through Technological Transfer,
  I$^2$T$^3$ Onlus\\
V.le Morgagni 67/a \\
50134 Firenze\\
E-mail: {\tt alessandro.speranza@i2t3.unifi.it}\\
A. Terenzi\\
Snamprogetti S.p.A.\\
Via Toniolo 1\\
61032 Fano (PU)\\
E-mail: {\tt alessandro.terenzi@snamprogetti.eni.it}
}
\maketitle
\abstract{
We propose a model for the simulation of the blowdown of vessels containing two-phase
(gas-liquid) hydrocarbon fluids, considering non equilibrium between phases. 
Two phases may be present either already at the beginning of the blowdown
process (for instance in gas-liquid separators) or as the liquid is formed
from flashing of the vapor due to the cooling induced by pressure decrease.
There is experimental evidence that the assumption of thermodynamic
equilibrium is not appropriate, since the two phases show an independent
temperature evolution. Thus, due to the greater heat transfer between the liquid
phase with the wall, the wall in contact with the
liquid experiences a stronger cooling than the wall in contact with
the gas, during the blowdown. As
a consequence, the
vessel should be designed for a lower temperature than if it was supposed
to contain vapor only.\\
Our model is based on a compositional approach, and it takes into account
internal heat and mass transfer processes, as well as heat transfer with the
vessel wall and the external environment. \\
Numerical simulations show a generally good agreement with experimental
measurements. 
}

\section{Introduction}

The problems related to the blowdown of pressure vessels containing mixtures of hydrocarbons
are well known amongst industries involved in plant designing and
hydrocarbons extraction. In particular, the pressure and thermal
stress to which the vessel is exposed during the blowdown can
present a number of consequences such as cracks in the walls, that
have to be predicted by an accurate simulation of the blowdown
process. Problems are highly enhanced when some liquid is
present in the vessel either already at the start of the
depressurisation, or as a result of
condensation of heavier hydrocarbons, induced by the cooling caused by
the expansion. Clearly, being the depressurisation usually caused by
security alerts, the blowdown process is often very rapid. Typically
the pressure can fall by 100 bar in a few hundreds 
seconds. Depending on the concentration of the mixture of
hydrocarbons, the fluid temperature can fall of as much as 100 K in
the same time. In the case of presence of condensed liquid, the vessel
wall in contact with it, as an effect of the higher thermal conductivity
of the liquid, compared to the one of the gas, can be exposed to a
temperature drop of almost as much as the liquid itself, \ie, 50-100
K in a few minutes~\cite{HaqRicSavCha92}. With these extreme working
conditions, the modeling of the blowdown process have to be made by
taking into account as many factors as possible. These 
include heat transfer with the external environment, the presence of
many components in the vessel and the possibility of situations in
which the assumption of phase equilibrium are not appropriate.
In fact, in particular this aspect, which is comforted by experimental
evidence~\cite{HaqRicSavCha92}, is usually not considered in thermodynamic
commercial programs. Furthermore, the complexity of the thermodynamic
behavior caused by the presence of 
many different hydrocarbons, rather than one pure fluid, often induce
authors to neglect it in first approximation~\cite{XiaSmiYad93}. As far as we know, the only 
complete model analysed in the past is the one described by Haque {\em
  et al.}~\cite{HaqRicSav92}. There, the authors claim their model to allow for
a multicomponent system, non-equilibrium conditions between the gas
and the liquid within the vessel, and possibility of having separate
water as well as gas and liquid phase. 
Also all the heat transfers, except for the one between the two
fluids, are taken into account of. However, despite the numerical
results obtained by the authors show a good agreement with some
experiments~\cite{HaqRicSavCha92}, the model is completely
undocumented. A full understanding of the actual process and the
mathematics involved is therefore, at present, beyond reach.

For these reasons, and with the financial and technical support of
ENI S.p.A., a leading company involved in pipelines and fuels, we develop
a complete model taking into account of all the factors above. Being
interested in the evolution of the pressure within the vessel and the
temperature of the fluids and the walls in contact with them, we
neglect all the spatial variations of the thermodynamic quantities and
concentrate on their evolution in time. The assumption of homogeneous\ fluid
is consistent with what is actually measured experimentally during the
blowdown, and therefore a priori justified by the necessity to focus
on average quantities, rather than on their local
variations. Furthermore,  even a rapid blowdown process, 
occurs at a time-scale which is generally much longer than the time
needed for the pressure within the vessel to rearrange. 
Therefore, local variations of the pressure should
generally be neglected in a vessel. This is not as clear for the rearrangement of the
temperature within the fluids. However, given the violence of the
pressure drop, especially in the early stages of the depressurisation,
the liquid, if present, will quickly start boiling strongly. This
justifies, through a strong convection, the homogeneity of its temperature. For
the gas, on the other hand, assuming a rapid motion induced by the acceleration
of the gas far upstream the orifice towards the exit, we can imagine
it to get mixed and homogenized at all the time, especially in the
early stages of the blowdown,  while the pressure is dropping steeply.

Our model is only partially based on the description of the model by
Haque et al. in~\cite{HaqRicSav92}. In fact, it differs from it
rather deeply in the treatment reserved to the energy
balance. Furthermore, we neglect the presence of separate water and
take into account the heat exchanged at the interface between the
fluids. Here we underline the assumptions made, the basic equations, and the
numerical scheme used to solve the model.
In order to account for the mass exchanged between the gas and the
liquid phase, in conditions away from the thermodynamic equilibrium,
we introduce the ``partial phase equilibrium assumption'' that we
describe in Sec.~\ref{sec:model}.

\section{The model for mixtures of hydrocarbons}\label{sec:model}
As we mentioned in the introduction, in spite of the observed the non-equilibrium
conditions between the two phases (gas and liquid), possibly present in
the vessels during the blowdown process, we have mass exchange between
them. This means we have phase transition in one or both
senses. In reality, being a globally non-equilibrium problem, it will be
just the material at the interface between the two phases actually
exhibiting the phase transition in either senses. However, since, as
we mentioned, we neglect spatial variations of temperature and
pressure, we will have to assume some kind of phase equilibrium
between the two phases, in order to solve the problem of mass
exchange. In order to do this, besides the two bulk phases gas ``G''
and liquid ``L'', we will introduce two intermediate phases, present
instantly only in traces in the vessel, which are responsible for the phase
transition between the two bulk phases. We will call these two phases,
drops ``l'', \ie, the liquid that condenses from the gas and that,
immediately after condensation falls into the bulk liquid, and vapor
``g'', which is the gas that vaporizes from the bulk liquid phase, and
that migrates into the bulk gas region, immediately after forming.

Since the two bulk phases are in non-equilibrium conditions, \ie, at
different temperatures $\temp\G$ and $\temp\L$, while the two
incipient phases ``l'' and ``g'' has to be in equilibrium with their
respective parent phases ``G'' and ``L'', we will assume
$\temp\l=\temp\G$ and $\temp\g=\temp\L$. Thus, as the
incipient phases move into their respective bulk phases (g moves into
L and l moves into G) they slightly influence their temperatures, as
they mix with them. On the other hand, the two incipient phases mix
with the two bulk phases, and thus, they homogenize with them.

We will call the assumption above, ``partial phase equilibrium'', since we
assume that the two incipient (daughter) phases are instantly in
equilibrium with their respective parents. However, as soon as they
move into their bulk phases, they mix with them and disappear. This
process is repeated all the time, as long as phase transition occurs,
in either sense. Clearly, not necessarily the transition will 
be present in both senses. In fact, likely 
only vaporization will occur in most of the cases. Then, only ``g''
will be present in the vessel. However, as
temperature drops due to expansion, the heavier
components of the gas might condense, making the phase transition run in
both directions.

Let us then proceed to the underline of the full model for a
multicomponent system. Since it would lead us beyond the scope of the 
present work, here we will not derive the form of basic equations of
the model. However, for a clear derivation and general considerations
on the basic assumptions made, we will refer the reader
to good descriptions of the experimental setting and the basic model,
by Dutton and co-workers~\cite{DutCov97,Dutton99}.

\subsection{Polydisperse mass and energy balance}
Our model, is based on the mass and energy balance equations, between
the two bulk phases G and L. Since the two daughter phases appear and
disappear all the time, and are instantly present just in traces in
the vessel, we will not consider them in the global balance, but they
will appear as source terms for the balance equations.

Let us introduce $\mol_i\ph$, the number of moles of component $i$ in each phase (G and
L) and $\src_i\out$, the function describing the discharge~\cite{Blevins84} of
component $i$. Let us also introduce
$\src_i\vap$ and $\src_i\cond$, 
as the functions describing the rate of vaporization (\ie\ the number
of moles of component $i$ moving from L to g per unit time) and
condensation (\ie\ the number of moles of component $i$ passing
from G to l per unit time) of component $i$, respectively.

Since the internal energy and enthalpy are written for the
phase, and not for the single species, we will write just two energy
balance equations, while we will write one equation for each species
and each phase (G and L) for the number of moles (mass balance).
Assuming we have M components, we will therefore  have
\bea
\frac{d\mol_i\G}{dt}&=&-\src_i\out+\src_i\vap-\src_i\cond\label{eq:poly_moles_G} \\
\frac{d\mol_i\L}{dt}&=&-\src_i\vap+\src_i\cond\label{eq:poly_moles_L}
\eea
\[
{\rm for}\quad i=1\ldots {\rm M}
\]
From the 2M equations above we will get the evolution of $\mol_i\G$
and $\mol_i\L$ due to the discharge and the net balance between
condensation and vaporization.

The energy balance is, on the other hand, simply
\bea
\frac{d}{dt}\left(\mol\G\inte\G\right)&=&-\ent\G\src\out-\ent\G\src\cond-\left(\ent\l-\ent\G\right)\src\cond+\ent\g\src\vap\nonumber\\
&&+\hcoeff\LG
S\LG(\temp\L-\temp\G)+\hcoeff_{\rm WG}S_{\rm WG}(\temp\WG-\temp\G)\label{eq:poly_en_gas}\\
\frac{d}{dt}\left(\mol\L\inte\L\right)&=&\ent\l\src\cond-\ent\L\src\vap-\left(\ent\g-\ent\L\right)\src\vap\nonumber\\
&&-\hcoeff\LG
S\LG(\temp\L-\temp\G)+\hcoeff_{\rm WL}S_{\rm WL}(\temp\WL-\temp\L)\label{eq:poly_en_liq}
\eea
where the global quantities are simply obtained by summing the single species
quantities; for instance, $\src\vap=\sum_i\src_i\vap$. Here $S\LG$ is
just the area of the interface between the two phases and $S_{\rm WL}$
and $S_{\rm WG}$ is the area of interface between the phases and the
wall in contact with them, while the $\lambda$ coefficients are just
the thermal exchange coefficients between the two phases and the
phases and the walls in contact with them~\cite{Hollands84,Ford55,ElyHan81,ElyHan83}.

To the 2M+2 above equations, we will add the obvious total volume
conservation
\beq\label{eq:vtot_poly}
V_0=\mol\G\vol\G+\mol\L\vol\L
\eeq
and the four equations of state, one for each phase (see below), 
for a total of 2M+7 equations in the
2M+7 unknowns $\mol_1\G\ldots\mol_{\text{M}}\G,\
\mol_1\L\ldots\mol_{\text{M}}\L$, $\vol\G,\ \vol\L,\ \vol\l,\ \vol\g,\
\pres,\ \temp\G,\ \temp\L$. The two quantities $\temp\WL$ and
$\temp\WG$ are obtained by solving the two Fourier equations across
the wall, coupled with the system above.

\subsection{Polydisperse phase equilibrium}
Since the hydrocarbons have different molecular weight, 
the rate of condensation and vaporization will be different from
species to species, as different different is their tendency to be in the
gas or liquid phase (expressed in formal terms by their chemical
potentials). 
Different will also be the discharge rate,
$\src_i\out$, since as we can imagine, generally the lighter
components will tend to stay in the top part of the vessel and thus
closer to the orifice. Especially in the early stages of the 
blowdown, we can therefore imagine the gas phase to lose part of its
lighter components. However, we will neglect this aspect and we will assume
simply that $\src_i\out=\molfrac_i\G\src\out$, where 
$\molfrac_i\ph=\mol_i\ph/\sum_j\mol_j\ph$ is the mole fraction of
species $i$ in the phase $a$ (the sum is extended to M) and
$\src\out$ is the total discharge rate. In other words, we will assume
the gas to be perfectly mixed, assumption which is actually
consistent with the neglect of spatial variation of $\pres$ and $\temp$.

The values of $\src_i\cond$ and $\src_i\vap$ will be then obtained by
solving the phase equilibrium conditions for the two sub-systems G$\leftrightarrow$l and
L$\leftrightarrow$g, in which, in fact, we have split our
system. Since we now have a multicomponent system, we will need to solve
\[
\begin{array}{l|l}
\mu_i\G=\mu_i\l &\mu_i\L=\mu_i\g \\
\end{array}
\]
\[
{\rm for}\quad i=1\ldots {\rm M}
\]
From the solution of the phase equilibrium conditions above, we will
get $\mol_i\l$ and $\mol_i\g$ as functions of $\pres,\ \temp\G,\
\temp\L$ and the composition of the two parent
phases\footnote{\selectfont{Note here that the parent depends on time,
    since there is some fractionation which is induced by the discharge of gas
  only. Since presumably the most volatile hydrocarbons will mainly be
in gas phase, the discharge of gas only will leave only the heavier
hydrocarbons in the vessel.}}\ ($\mol_i\L$ and $\mol_i\G$). At fixed $t$ and thus at fixed
$\pres(t)$ and $\temp\G(t)$ and $\temp\L(t)$, we will therefore need
to solve the above phase equilibria to get $\mol_i\g,\ \mol_i\l$. The
rate of condensation and vaporization are  thus:
\bea
\src_i\cond&=&\frac{d\mol_i\l}{dt}\label{eq:poly_psicond}\\
\src_i\vap&=&\frac{d\mol_i\g}{dt}\label{eq:poly_psivap}
\eea

\section{Numerical results}

Here we will present just the results obtained with our model,
relatively to two experiments (I1 and S9) reported
in~\cite{HaqRicSavCha92}. However, these are just two preliminary
results, while we will apply our model to far more complex
environments. In any case, it is useful to use our model in these two 
particular situations, since there Haque and
co-workers describe the experimental apparatus and give some
experimental results that can be directly compared with what we obtain
from our model.

%
%
\subsection{Experiment I1}
The experiment consists on the blowdown of a cylindrical steel vessel,
with base diameter 0.273 m, length 1.524 m, wall thickness 2.5 cm and
top choke of diameter 0.635 cm. The vessel contains only N$_2$ at 150
bar and 290.15 K. The vessel is immersed in stagnant air at 290.15 K,
thus everything is in equilibrium before opening the valve.

In Fig.~\ref{fig:I1_pres}(a) we report the value of the pressure inside
the vessel during the blowdown, predicted and experimental. The
agreement is rather good, besides a small deviation from the
experimental data, roughly at half
process. As can easily be seen on a logarithmic scale, the blowdown is
almost exactly exponential. This is hardly surprising, being the
conditions inside the vessel comparable to the ideal gas.
\bfig
\begin{picture}(0,0)%
\includegraphics{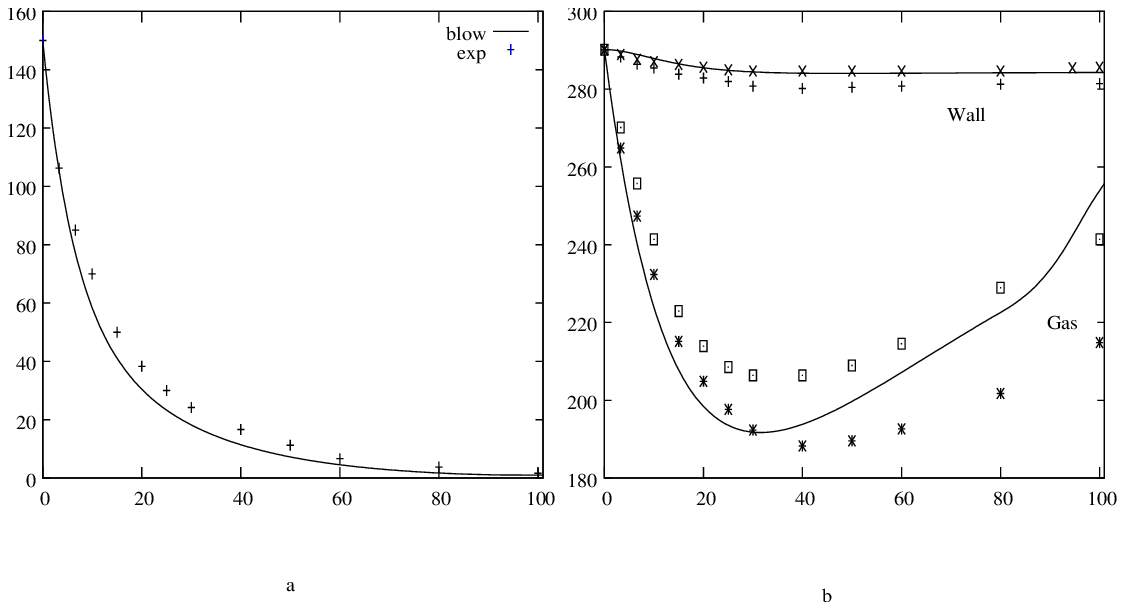}%
\end{picture}%
\setlength{\unitlength}{2763sp}%
\begingroup\makeatletter\ifx\SetFigFont\undefined%
\gdef\SetFigFont#1#2#3#4#5{%
  \reset@font\fontsize{#1}{#2pt}%
  \fontfamily{#3}\fontseries{#4}\fontshape{#5}%
  \selectfont}%
\fi\endgroup%
\begin{picture}(7791,4064)(1043,-4561)
\put(6755,-4336){\makebox(0,0)[lb]{\smash{{\SetFigFont{6}{7.2}{\familydefault}{\mddefault}{\updefault}{\color[rgb]{0,0,0}$t$ (s)}%
}}}}
\put(5049,-2461){\rotatebox{90.0}{\makebox(0,0)[lb]{\smash{{\SetFigFont{6}{7.2}{\familydefault}{\mddefault}{\updefault}{\color[rgb]{0,0,0}$\temp$ (K)}%
}}}}}
\put(3049,-4186){\makebox(0,0)[lb]{\smash{{\SetFigFont{6}{7.2}{\familydefault}{\mddefault}{\updefault}{\color[rgb]{0,0,0}$t$ (s)}%
}}}}
\put(1149,-2386){\rotatebox{90.0}{\makebox(0,0)[lb]{\smash{{\SetFigFont{6}{7.2}{\familydefault}{\mddefault}{\updefault}{\color[rgb]{0,0,0}$\pres$ (bara)}%
}}}}}
\end{picture}%
\caption{a):Pressure behavior during experiment I1 obtained by our model
  {\em blow}  and experimental  results  from Haque and co-workers. The
  agreement is rather good.
  b):Temperature of the gas and the wall in contact with it,
  obtained by our model, compared with the experimental
  results. Experimental results are contained between the two
  discontinuous curves. Again our numerical results lie between the
  two experimental curves.}
\label{fig:I1_pres}
\label{fig:I1_all_temp}
\efig

In Fig.~\ref{fig:I1_all_temp}(b) we show the behavior of the temperature
of the gas inside the vessel, and the internal wall in contact with it,
obtained with our model, compared to the experimental results reported
in~\cite{HaqRicSavCha92}. Once again our result lies within the
region spanned by the two experimental curves (the two discontinuous curves) and thus
in excellent agreement with them. The plot in
Fig.~\ref{fig:I1_all_temp}(b) shows that the temperature is driven by
the cooling due to the expansion in the first stage of the
depressurisation, up to c.ca 30 s. Later the heat coming in from the
outside becomes important and the gas is heated up by the walls. For
large $t$, the temperature goes back to equilibrium with the
external environment and the pressure stops dropping.

The numerical results obtained by our model, of the temperature
evolution of the wall surface in contact with the gas (see
Fig.~\ref{fig:I1_all_temp}b), are again in excellent agreement with the
experimental data, as our prediction lies between the two experimental curves.

\subsection{Experiment S9}

In this experiment, we model a cylindrical vessel of 1.130 m diameter,
3.240 m length, 5.9 cm wall thickness and top choke of 1 cm diameter.
The initial composition is of 85.5 mole \% methane, 4.5 mole \% ethane,
10.0 mole \% propane, at 120 bara and 290.15 K. The vessel
is immersed in stagnant air at 290.15 K. 

The pressure evolution predicted by our model is reported in
Fig.~\ref{fig:S9_pres}(a). 
\bfig
\begin{picture}(0,0)%
\includegraphics{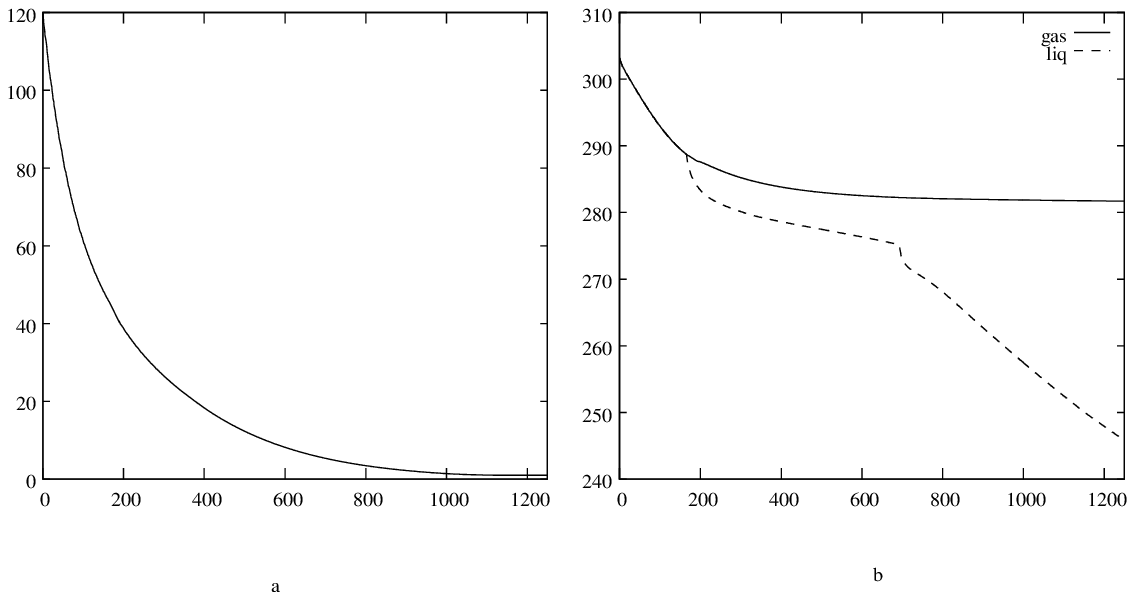}%
\end{picture}%
\setlength{\unitlength}{2763sp}%
\begingroup\makeatletter\ifx\SetFigFont\undefined%
\gdef\SetFigFont#1#2#3#4#5{%
  \reset@font\fontsize{#1}{#2pt}%
  \fontfamily{#3}\fontseries{#4}\fontshape{#5}%
  \selectfont}%
\fi\endgroup%
\begin{picture}(7826,4004)(1115,-4486)
\put(5132,-2611){\rotatebox{90.0}{\makebox(0,0)[lb]{\smash{{\SetFigFont{6}{7.2}{\familydefault}{\mddefault}{\updefault}{\color[rgb]{0,0,0}$\temp (K)$}%
}}}}}
\put(7101,-4261){\makebox(0,0)[lb]{\smash{{\SetFigFont{6}{7.2}{\familydefault}{\mddefault}{\updefault}{\color[rgb]{0,0,0}$t (s)$}%
}}}}
\put(2995,-4261){\makebox(0,0)[lb]{\smash{{\SetFigFont{6}{7.2}{\familydefault}{\mddefault}{\updefault}{\color[rgb]{0,0,0}$t$ (s)}%
}}}}
\put(1225,-2536){\rotatebox{90.0}{\makebox(0,0)[lb]{\smash{{\SetFigFont{6}{7.2}{\familydefault}{\mddefault}{\updefault}{\color[rgb]{0,0,0}$\pres$ (bara)}%
}}}}}
\end{picture}%
\caption{a): Pressure evolution of experiment S9 against time. The
  pressure decrease is again almost exponential and rather close to
  the results reported in the paper by Haque and co-workers. b):
  $\temp$ evolution of the wall in contact with the two fluids. After
  the liquid appears (see text), the wall in contact with it cools
  much more rapidly than the wall in contact with the gas as is
  expected.}
\label{fig:S9_pres}
\label{fig:S9_wall}
\vspace*{-0.3cm}
\efig
As we can observe from Fig.~\ref{fig:S9_pres}(a), the pressure drop is
rather slower than in experiment I1. This is probably due to the less
efficient expansion of the heavier hydrocarbons in
the parent, and the lower value of the initial pressure. As a
consequence, the temperature of the wall in contact with the gas (see
below), although slowly, drops rather more significantly than in the
previous example.  

In Fig.~\ref{fig:S9_temp_PT}(a), we show the temperature evolution of the
fluid inside the vessel.
\bfig
\vspace*{-0.4cm}
\begin{picture}(0,0)%
\includegraphics{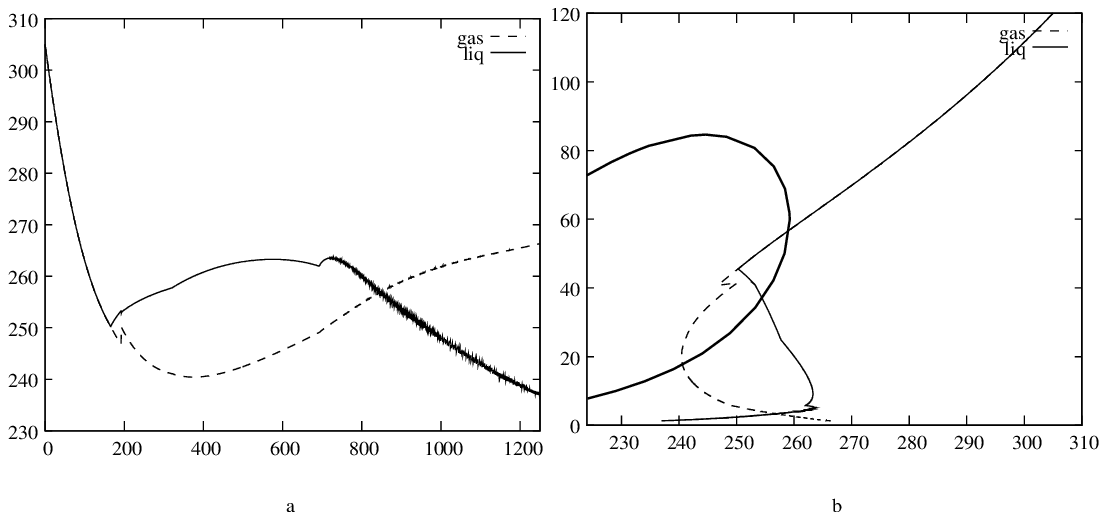}%
\end{picture}%
\setlength{\unitlength}{2763sp}%
\begingroup\makeatletter\ifx\SetFigFont\undefined%
\gdef\SetFigFont#1#2#3#4#5{%
  \reset@font\fontsize{#1}{#2pt}%
  \fontfamily{#3}\fontseries{#4}\fontshape{#5}%
  \selectfont}%
\fi\endgroup%
\begin{picture}(7659,3445)(1062,-3886)
\put(6734,-3736){\makebox(0,0)[lb]{\smash{{\SetFigFont{6}{7.2}{\familydefault}{\mddefault}{\updefault}{\color[rgb]{0,0,0}$\temp (K)$}%
}}}}
\put(5040,-2236){\rotatebox{90.0}{\makebox(0,0)[lb]{\smash{{\SetFigFont{6}{7.2}{\familydefault}{\mddefault}{\updefault}{\color[rgb]{0,0,0}$\pres$ (bar)}%
}}}}}
\put(1160,-2086){\rotatebox{90.0}{\makebox(0,0)[lb]{\smash{{\SetFigFont{6}{7.2}{\familydefault}{\mddefault}{\updefault}{\color[rgb]{0,0,0}$\temp (K)$}%
}}}}}
\put(3073,-3661){\makebox(0,0)[lb]{\smash{{\SetFigFont{6}{7.2}{\familydefault}{\mddefault}{\updefault}{\color[rgb]{0,0,0}$t (s)$}%
}}}}
\end{picture}%
\caption{(a) -- Evolution of the temperature of the fluid in experiment
  S9. Liquid appears after about 165 s from the start of the
  depressurisation. The liquid is
  immediately heated up by the wall, due to its larger thermal
  conductivity, compared to the one of the gas. For large $t$, after
  the depressurisation is complete, the two
  fluids tend to have the same temperature.\newline
(b) -- $\pres$ against $\temp$ evolution of experiment S9, together
  with the two phase coexistence region. During the blowdown, the system enters the phase
  coexistence region, but only after some time, the liquid (dashed
  line) actually appears. Clearly, being the compositions of the two phases
  different and different from the parent, after that point, the phase
  envelope does not correspond to the phase envelope of either phase.
}
\label{fig:S9_temp_PT}
\vspace*{-0.5cm}
\efig
Liquid appears after about 165 s from the start of the blowdown, with
some delay, after the crossing of the phase boundary. This delay is
probably due to the rapidity of the blowdown process, that makes the
phase transition occur in the ``spinodal region'', \ie, where the parent
becomes unstable~\cite{SolWarCat01}, rather than at the ``cloud
point'', \ie, where the two phases actually start to coexist.
However, this delay, although confirmed by the
experimental results reported in~\cite{HaqRicSavCha92}, could as well
be just a numerical effect. In fact, our program does not include any
stability check of the free energy and the phase equilibrium
equations are only solved without looking for a globally more stable
solution~\cite{SolWarCat01}. It could well be that the liquid actually
appears earlier than we found with our model. However, this aspect,
although interesting, lies beyond our present scope. 

Note the jump of the gas temperature after about 200 s, immediately after the
appearance of the liquid phase. This abrupt increase in the
temperature is due to the heat of condensation transmitted to the gas
by a finite quantity of material passing into the liquid
phase. Correspondingly, there is a small decrease of the liquid
temperature, due to the influence of this quantity of drops at
temperature $\temp\G < \temp\L$ falling into the liquid bulk phase. 
Note also the appearance of some numerical noise in later stages. These
oscillations are probably due to some errors in the evaluation of the
internal energy of the liquid, when the quantity of liquid becomes
small due to vaporization. The temperature of the liquid appears to be
rather sensitive to changes in the value of the internal energy.    

The delay after which the liquid phase appears,
is more evident when we plot the $\pres$ against
$\temp$ evolution, together with the phase diagram relative to the
initial composition (see Fig.~\ref{fig:S9_temp_PT}(b)). The two phase
coexistence region is bordered by the phase envelope (thick
continuous curve in Fig.~\ref{fig:S9_temp_PT}(b)). The $\pres$
vs. $\temp$ evolution predicted by our model during the
depressurisation correctly cross the envelope (phase boundary) and
the liquid appears after some delay, well inside the coexistence region.

There is also to notice, that when the liquid solution is found, we do
not actually have an incipient liquid phase and a parent gas phase. In
fact, the first
solution with the two phases, already consists of a macroscopic
presence of liquid and therefore the cloud point is already far
behind and has actually been missed by our algorithm. 
It is possible that a more precise investigation of the onset
of the phase coexistence, \eg, by reversing and reducing the step to
look for the point at which the liquid actually appears, would lead us
closer to the phase envelope. However, again despite being
scientifically an interesting point, we decided to overlook this phenomenon, and
concentrate on the temperature evolution of the vessel walls.

In Fig.~\ref{fig:S9_wall}(b) we plot the temperature of the wall portions
in contact with the two fluids against time. As it can be seen, the
whole surface is at a certain uniform temperature when the liquid
appears, as it should be, since the vessel is up to that point filled
with gas only. Furthermore, since we apply ``partial phase
equilibrium assumption'', the liquid, when it's formed, is instantly in phase equilibrium
with the gas. However, due to the abrupt change of the thermal conductivity (higher
for the liquid), the liquid absorbs much more heat from the wall, and
the wall in contact with it cools much more rapidly than the wall in
contact with the gas. This result is expected, and confirmed by the
experiments in~\cite{HaqRicSavCha92}. On the other hand, the liquid is
immediately heated up by the wall and its temperature rises abruptly
after its formation, above the temperature of the gas.

Furthermore, the strong cooling of about 50 K experienced by the
wall in contact with the liquid is again excellent agreement with the
experimental and numerical results
shown in~\cite{HaqRicSavCha92}. Finally, at present, we have no
physical explanation for the sudden change in the
steepness of the temperature in contact with the liquid at about 700
s, and corresponding to a similar change in the temperature of the
liquid. This jump in the derivatives corresponds to a big discontinuity
of the viscosity of the liquid phase, that suddenly jumps down of two
orders of magnitude to become similar to the viscosity of the
gas. This is probably a non physical behavior, induced simply by the
model for the viscosity, which based on the description of the models
by Ely and Hanley~\cite{ElyHan81,ElyHan83}.  Further investigations are
being carried out. 

\section{Conclusions}

We presented a model for the blowdown of pressure vessels containing a mixture of
hydrocarbons. Our model is based on a global mass and
energy balance between the phases, gas and occasionally liquid, present
in the vessel, at every stage of the blowdown. 
The strong cooling to which the vessel wall may be exposed, especially in
presence of some liquid, requires an accurate modelling which takes
into the account as many phenomena as possible, in order to avoid
cracks in the vessel. Our model, which is based on a compositional
approach, allows for the presence of many
different hydrocarbons within the vessel, as well as non-equilibrium
conditions between the phases. All the heat exchanged between the
fluids and with the external environment, via heat diffusion across
the vessel walls are also considered. In order to account for the mass
and energy exchanged between the phases,
a partial phase equilibrium assumption has been made. Within this
approximation, we assume that the phase formed by condensation and/or
vaporization, remains instantly in equilibrium with its parent phase, before
moving into the corresponding bulk phase 
and homogenizing with it.  

The numerical results show a rather good agreement with some
experimental results, both qualitatively and quantitatively. In
particular, in the case of a mixture of methane, ethane and propane, the
appearance of a liquid phase after some time from the start of the
depressurisation is correctly predicted. The blowdown is estimated to
be complete in about 1200 s, in line with the experiments. Also the
temperature drop of the fluids and the wall in contact with them is
quite well estimated by our model. Similarly, for experiment I1,
consisting on the blowdown of a vessel containing N$_2$ only, the drop
of the pressure and temperature of the gas and of the wall in contact with it
are all in excellent agreement with the experimental results.

\section*{Acknowledgments}
AS wishes to thank Prof. A. Fasano and
Prof. M. Primicerio from Universit\`a degli Studi di Firenze for their advice,
ENI S.p.A. and I$^2$T$^3$ Onlus, for funding this research, Snamprogetti S.p.A.,
for providing technical and bibliographical support and
Enitecnologie in the person of Dr. D. Bersano, for providing the
subroutines upon which the phase equilibrium calculation is based.
\bibliographystyle{abbrv}

\end{document}